\documentclass[prb,aps,amssymb,twocolumn,showpacs,floatfix]{revtex4}

\usepackage{graphicx}

\begin{document}

\title{Spin Dynamics and Level Structure of Quantum-Dot Quantum Wells}

\author{Jesse Berezovsky}
\author{Min Ouyang}
\author{Florian Meier}
\author{David D. Awschalom}
\affiliation{Center for Spintronics and Quantum Computation,
University of California, Santa Barbara, California 93106, USA }
\author{David Battaglia}
\author{Xiaogang Peng}
\affiliation{Department of Chemistry and Biochemistry, University
of Arkansas, Fayetteville, Arkansas 72701, USA}
\date{\today}

\begin{abstract}
We have characterized CdS/CdSe/CdS quantum-dot quantum wells using
time-resolved Faraday rotation (TRFR). The spin dynamics
show that the electron $g$-factor varies as a function of quantum well width and the
transverse spin lifetime of several nano-seconds is robust up to room temperature.
As a function of probe
energy, the amplitude of the TRFR signal shows pronounced
resonances, which allow one to identify individual exciton
transitions. The resonance energies in the TRFR data are consistent with different
exciton transitions in which the electron occupies the conduction band ground
state.
\end{abstract}


\maketitle

Nanocrystals have promising applications in optics and spin- or
charge-based quantum information schemes because electrons are
confined on a nanometer scale. The implementation of quantum
information schemes would require several nanocrystals to be
assembled into functional structures. For nanocrystals
interconnected by conjugated molecules, spin-conserving electron
transfer  between nanocrystals has been
demonstrated.~\cite{ouyang:03} Quantum-dot quantum well (QDQW)
heterostructures, where layers of different semiconducting
materials alternate in a single nanocrystal, represent an
alternative pathway towards the synthesis of functional
structures. Both core-shell quantum
dots~\cite{hines:96,dabbousi:97,li:03}  and
QDQWs~\cite{kortan:90,mews:94,mews:96,tian:96,little:01,battaglia:03}
have been synthesized during the past years. QDQWs with a
large-bandgap core allow one to investigate quantum confined
levels in a geometry in which electrons occupy the surface of a
sphere. Both
CdS/HgS/CdS~\cite{mews:94,schooss:94,mews:96,jaskolski:98,bryant:03}
and CdS/CdSe/CdS~\cite{battaglia:03} QDQWs have been well
characterized by photoluminescence (PL) and absorption
spectroscopy. However, a detailed investigation of the quantum
size levels is challenging because of inhomogeneous broadening.
Individual exciton transitions have so far only been resolved with
techniques such as hole burning, where a subset of homogeneous particles is selected
spectroscopically.~\cite{mews:96} The electron spin dynamics in
QDQWs have not yet been addressed.

Here, we report time-resolved Faraday rotation
(TRFR)~\cite{baumberg:94,gupta:02} for CdS/CdSe/CdS QDQWs with
varying CdSe quantum well width ($n_{\rm CdSe}=1-5$ monolayers).
The spin lifetime is of order $2 - 3 \,{\rm ns}$ and almost
temperature-independent up to $294 \, {\rm K}$, comparable to CdSe
quantum dots.~\cite{gupta:99} The QDQWs exhibit $g$-factors that
vary with quantum well width. TRFR is not only a unique
experimental probe for the spin dynamics, but also a sensitive
spectroscopic technique. In contrast to absorption spectra, the
amplitude of the TRFR signal as a function of probe energy
exhibits several distinct resonances close to the absorption edge,
because optical transitions to the lowest conduction band level
are probed selectively. From the level scheme and dielectric
response functions evaluated with ${\bf k}\cdot {\bf p}$
calculations,~\cite{jaskolski:98,pokatilov:01} we show that the
resonance energies in the TRFR data are consistent with the
conduction and valence band level scheme of spherical QDQWs. In
contrast, the spectral weight of the resonances is not correctly
reproduced.

\begin{figure}[!t]
\centerline{\mbox{\includegraphics[width=7cm]{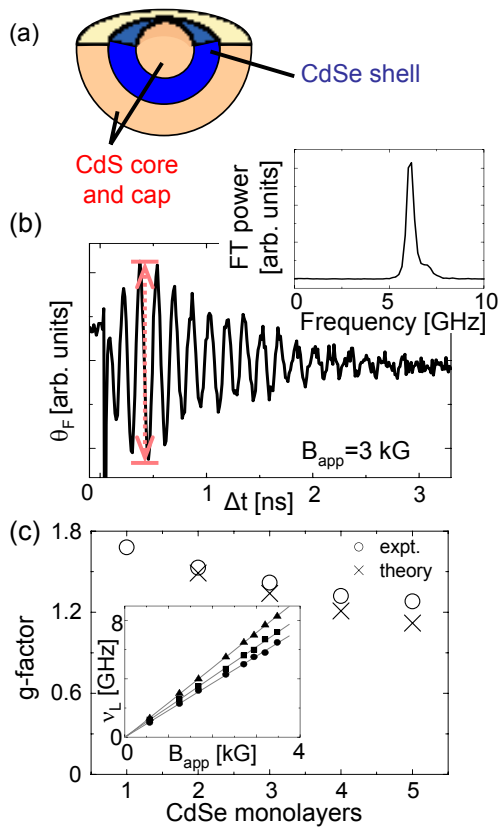}}}
\caption{(color online). (a) Schematic representation of the QDQW. (b)
Typical TRFR data from a QDQW with $n_{\rm CdSe}=3$ and $B_{\rm app}=3 \, {\rm kG}$.
The dotted arrow indicates how the amplitude of $\theta_F$ is determined for
Fig.~\ref{Fig2exp}. Inset: FT power spectrum of the data. (c) Electron $g$-factor as a function of
 CdSe quantum well width.  The measured values (circles) are compared to calculated $g$-factors
 (crosses).  The second $g$-factor with smaller amplitude is not shown.  Inset:
$\nu_L$ as a function of $B_{\rm app}$ for $n_{\rm CdSe} = $1 (triangles),
 3 (squares), and 5 (circles).
}\label{Fig1exp}
\end{figure}

{\it Experimental results.--} Colloidal QDQWs with varying width
of the CdSe quantum well were
synthesized by a successive ion layer adsorption and reaction
(SILAR) technique to produce nanocrystals with
accurate control over the quantum well width.~\cite{li:03,battaglia:03} A
schematic representation of the structure is shown in
Fig.~\ref{Fig1exp}(a). The QDQWs were dissolved in toluene and
all measurements were carried out in solution at $294 \, {\rm K}$
unless otherwise specified.

A regeneratively amplified Ti:Sapphire laser was used to generate
pump and probe pulses of independently tunable wavelength and
$\sim 200 \, {\rm fs}$ duration through optical parametric
amplification. In these measurements, the pump wavelength was
fixed at $\lambda_{\rm pump} = 505 \,{\rm nm}$. The pump and probe
pulses were both focused to a spot with a diameter of order $100 \, \mu{\rm m}$
within the QDQW solution. Spin-polarized electrons were excited into the conduction band
states of the QDQWs by the circularly polarized pump pulse.
Relaxation of the electron and hole to the lowest exciton state
presumably occurs on a picosecond time-scale, as in similar
systems such as CdS/HgS/CdS QDQWs.~\cite{braun:01} The linearly
polarized probe pulse then passes through the QDQW solution a time $\Delta t$ later,
where $\Delta t$ is set using a mechanical delay line in the pump beam path.  The
Faraday effect causes the polarization of the probe pulse
to be rotated by an angle, $\theta_F$, proportional to the component
of the net spin polarization along the probe beam direction.
By recording $\theta_F$ for varying $\Delta  t$, we detect the time
evolution of the optically injected electron spins in the QDQWs.

Two permanent magnets with adjustable separation were used to
apply a magnetic field, $B_{\rm app}$, to the sample perpendicular to the
pump and probe direction.  Spins that were initially polarized along the
pump beam precess around the magnetic field
at the Larmor frequency, $\nu_L = g \mu_B B_{\rm app}/h$ where $g$ is the
electron $g$-factor,
$\mu_B$ the Bohr magneton, and $h$ the Planck constant.  Figure~\ref{Fig1exp}(b)
shows typical data from a sample with a quantum well width of $n_{\rm CdSe}=3$ monolayers
and $B_{\rm app}= 0.3 \,{\rm T}$.  The inset shows the Fourier transform
(FT) power spectrum of the time-domain data.  A second
precession frequency was observed, as indicated both by the small shoulder in the FT
spectrum and the beating in the time-resolved data.  While the
origin of this second frequency is unclear in the present case, similar behavior
has been observed in CdSe nanocrystals.~\cite{gupta:02,rodina:03,schrier:03,chen:04}
There is also a non-oscillating
component to the TRFR signal which was also seen in previous
measurements on CdSe nanocrystals.~\cite{gupta:02}  In some samples, particularly for $n_{\rm CdSe}=5$,
the magnitude of the non-oscillating component
is comparable to that of the oscillating component.  However,
for the purposes of this paper we focus only on the oscillating
component [indicated by the arrow in Fig.~\ref{Fig1exp}(b)].  The effective
transverse spin lifetime, $T_2^\ast$, was
of order $2$ or $3 \,{\rm ns}$ for all samples measured.  The
spin lifetime was essentially temperature-independent between
room temperature and $5 \,{\rm K}$.~\cite{rem2}

We have performed TRFR measurements as a function of $B_{\rm app}$ on
samples with CdSe quantum well widths of $n_{\rm CdSe}=1 - 5$ monolayers.  In all
cases, the results show either one or two precession frequencies
that increase linearly with $B_{\rm app}$.   The inset of Fig.~\ref{Fig1exp}(c) shows the
main precession frequency as a function of $B_{\rm app}$ for $n_{\rm CdSe}=1, 3,$
and $5$ monolayers.  The measured $g$-factor for each sample is
shown in Figure~\ref{Fig1exp}(c) (circles) in comparison with the theoretical values (crosses)
obtained from an weighted average of the CdSe and CdS $g$-factors (see below). Because of the
fairly good agreement, we attribute the observed precession to the electron spin.
Within the experimental error, the $g$-factor did not show any dependence on
temperature from 5 K to room temperature~\cite{rem2} or on the probe
wavelength.

\begin{figure}[!t]
\centerline{\mbox{\includegraphics[width=6.5cm]{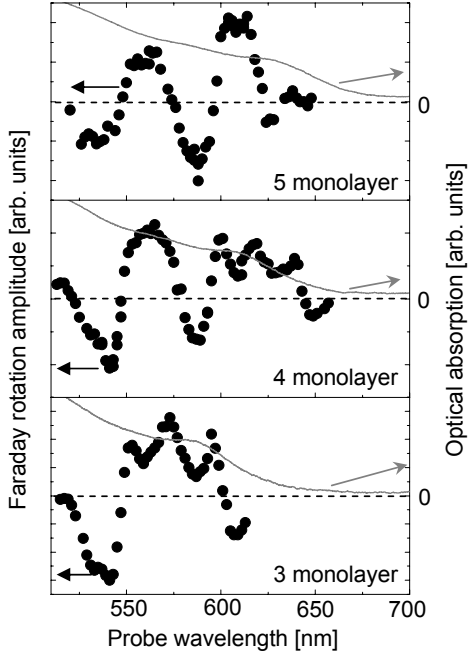}}}
\caption{Amplitude of the Faraday rotation angle, $\theta_F$, as a function
of probe wavelength for $n_{\rm CdSe}=3, 4,$ and $5$.  The numerical value
for $\theta_F$ was defined as the difference between the local maximum and minimum
of the oscillations in the TRFR data nearest to $\Delta  t=500 \, {\rm ps}$ [Fig.~\ref{Fig1exp}(b)], normalized
by the probe power.  The optical absorption for each sample is also shown.}\label{Fig2exp}
\end{figure}

In order to investigate the QDQW energy levels, we
have measured the dependence of the TRFR amplitude on probe wavelength
in the samples with $n_{\rm CdSe}=3, 4,$ and $5$.
The probe beam, which had a full-width at half maximum of $\sim 10 \, {\rm nm}$, was
passed through a monochromator after the sample yielding a wavelength
resolution of $2 \, {\rm nm}$.  Figure~\ref{Fig2exp} shows the
TRFR oscillation amplitude as a function of probe wavelength for the different samples together
with optical absorption data. While the
absorption signal only shows a featureless staircase-like behavior with no
distinct resonances, the amplitude of the TRFR signal exhibits several
pronounced resonances close to the absorption edge. The results in Fig.~\ref{Fig2exp}
show that TRFR not only provides information on the spin dynamics,
but also is a more sensitive spectroscopic technique than absorption spectroscopy
and allows one to identify individual exciton transitions in QDQWs.

{\it Theoretical description.--} We next turn to the theoretical
description of the experimental data. The conduction and valence
band level scheme of spherical QDQWs is calculated with ${\bf
k}\cdot {\bf p}$ theory,~\cite{jaskolski:98,pokatilov:01} using a
two-band description for the conduction band and the four-band
Luttinger Hamiltonian in the spherical approximation for the
valence band. The conduction band masses and Luttinger parameters
for CdSe and CdS are $\{m_{\rm CdSe}/m_0,\gamma_{1, {\rm
CdSe}},\gamma_{ {\rm CdSe}}\}=\{0.11,1.67,0.56\}$ and $\{m_{\rm
CdS}/m_0,\gamma_{1, {\rm CdS}},\gamma_{ {\rm
CdS}}\}=\{0.15,1.09,0.34\}$, respectively, where $m_0$ denotes the
free electron mass.~\cite{landolt,richard:96} We use the offset of
the CdS conduction and valence band edge relative to CdSe,
$0.32\,~{\rm eV}$ and $0.42\,~{\rm eV}$,~\cite{wei:00}
respectively, to define the radial potential for electrons and
holes. The inner and outer radius of the CdSe quantum well is
denoted by $r_1$ and $r_2$, respectively. The width of a CdSe
monolayer is approximated by the bulk value $0.43\,~{\rm nm}$
(Ref.~\onlinecite{li:04}) and the core radius and capping layer
width are $r_1=1.7\, {\rm nm}$ and $r_3-r_2=1.6\,{\rm nm}$,
respectively. Details are presented elsewhere.~\cite{meier:04b}

\begin{figure}[!t]
\centerline{\mbox{\includegraphics[width=7.3cm]{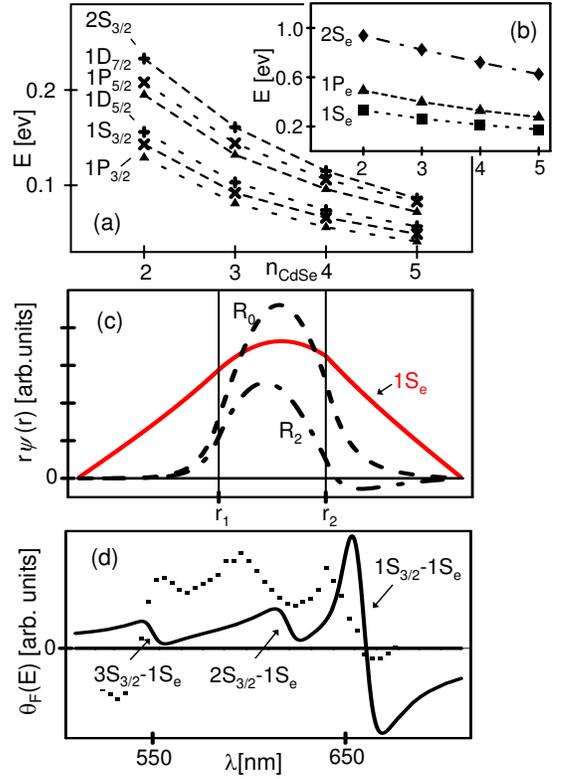}}}
\caption{(color online). (a) Lowest hole energy levels relative to the
CdSe valence band edge as a function of the quantum well width, $n_{\rm CdSe}$. (b) Conduction band energy
levels relative to the CdSe conduction band edge as a function of $n_{\rm CdSe}$.
(c) Radial wave function of the conduction band ground state
$1S_e$ (solid) and the $R_0$ (dashed) and $R_2$ (dashed-dotted)
components of the valence band state $1S_{3/2}$ for $n_{\rm
CdSe}=3$. (d) Amplitude of the TRFR signal, $\theta_F(E)$,
calculated from the level schemes in (a) and (b) for a spherical
QDQW with $n_{\rm CdSe} = 3$ and
$\gamma_v=15\, {\rm meV}$ (solid line) in comparison with experimental data (symbols).}\label{Fig1thy}
\end{figure}

The energies of the lowest conduction and valence band states are
shown in Figs.~\ref{Fig1thy}(a), (b). Different valence band
multiplets are denoted by $L_F$,~\cite{xia:89,efros:92} where $L$
is the smallest angular momentum of the envelope wave function and
$F$ the total angular momentum. Figure
~\ref{Fig1thy}(c) shows the radial wave function of the conduction
band ground state $1S_e$ (solid line) and of $1S_{3/2}$ (broken
lines) for $n_{\rm CdSe}=3$. Because of the larger valence band
mass, the valence band states are much better localized in the
quantum well.  The valence band ground state, $1P_{3/2}$, has a
$p$-type envelope wave function, which is consistent with a dark
exciton ground state.

From the energy $E_{1S_e}$ and wave function $\psi_{1S_e}({\bf r})$ of the
conduction band ground state $1S_e$, the electron $g$-factor is
estimated by an weighted average over the CdSe and CdS $g$-factors,
\begin{eqnarray}
g&=& g_{\rm CdSe} \int_{r_1}^{r_2} d {\bf r} \,  |\psi_{1S_e}({\bf r})|^2
\, \label{eq:g-av} \\
&& + g_{\rm CdS}\bigl( \int_{0}^{r_1} d {\bf r} \,  |\psi_{1S_e}({\bf r})|^2  +
\int_{r_2}^{r_3} d {\bf r} \,  |\psi_{1S_e}({\bf r})|^2 \bigr).
\nonumber
\end{eqnarray}
$g_{\rm CdSe}$ and  $g_{\rm CdS}$ are given by $g_{\rm CdSe/CdS}=2-
2  E_p \Delta_{\rm so}/3(E_g+\Delta_{\rm so}+E_{1S_e})(E_g+E_{1S_e})$,
where $E_p$, $E_g$, and $\Delta_{\rm so}$ denote the Kane
interband energy, bandgap, and spin-orbit energy of CdSe and
CdS, respectively. The energy of the conduction band ground state,
$E_{1S_e}$, is evaluated relative to the conduction band minimum.
Figure~\ref{Fig1exp}(c) shows the theoretical $g$-values (crosses)
obtained with standard parameters for $E_p$, $E_g$, and
$\Delta_{\rm so}$.~\cite{landolt} The agreement is
good for narrow QDQWs, but the theoretical value
Eq.~(\ref{eq:g-av}) is smaller than the experimental $g$-factor
for larger $n_{\rm CdSe}$. Possible explanations for this
discrepancy are the energy-dependence of the conduction band
mass~\cite{gupta:02} and interface terms in the expression for the
$g$-factor,~\cite{kiselev:98} which are neglected in
Eq.~(\ref{eq:g-av}).

From the calculated single-particle spectrum, we evaluate the amplitude of
the TRFR signal as a function of probe energy, $\theta_F(E)$,
which is proportional to the difference of the dynamic dielectric
response functions for $\sigma^\pm$ circularly polarized
light. The conduction band
electron with $s_z = 1/2$ created by the pump pulse relaxes
rapidly to $1S_e$, such that $\theta_F(E)$ is determined by
optical transitions to the unoccupied $1S_e$ state, $|1S_e;
\downarrow \rangle$,~\cite{hugonnard:94,linder:98,sham:99,meier:04}
\begin{eqnarray}
\theta_F(E)& = &C E \sum_{\sigma =\pm 1; |\Phi_v \rangle} \sigma
\left| \langle 1S_e; \downarrow |\widehat{p}_x + \sigma  i
\widehat{p}_y |\Phi_v \rangle \right|^2 \label{eq:frot} \\ &&
\hspace*{2cm}  \times \frac{E-E_{X,v}}{(E-E_{X,v})^2 + \gamma_v^2}. \nonumber
\end{eqnarray}
The sum extends over all valence band states $|\Phi_v \rangle$,
$E_{X,v}$ ($\gamma_v$) denotes the energy (linewidth) of the
$1S_e$-$\Phi_v$ exciton transition, and $C$ is a constant.
Equation~(\ref{eq:frot}) implies that only transitions to the {\it
conduction band ground state} contribute to $\theta_F(E)$. The
transition matrix element is finite for $S_{3/2}$ valence band
multiplets.~\cite{ekimov:85} Because the characteristic
energy splitting between these multiplets is of
order $0.1 \, {\rm eV}$, $\theta_F(E)$ exhibits several
well-defined resonances close to the absorption edge. If the
crystal anisotropy is taken into account,~\cite{efros:92} these
resonances split into doublets, but the characteristic energy
splitting is smaller than $25 \, {\rm meV}$. $\theta_F(E)$
exhibits distinct resonances for the $1S_{3/2}$, $2S_{3/2}$, and
$3S_{3/2}$ multiplets, with a spectral weight that is larger for
$1S_{3/2}$ than for $2S_{3/2}$ and $3S_{3/2}$ because of the
larger overlap with the envelope wave function of $1S_e$. For $n_{\rm CdSe}=3$,
$\theta_F(E)$ is shown in Fig.~\ref{Fig1thy}(d)  in comparison with
experimental data from Fig.~\ref{Fig2exp}. The energies of the
$1S_{3/2}-1S_e$, $2S_{3/2}-1S_e$, and $3S_{3/2}-1S_e$ transitions
are in good agreement with the experimental resonance energies.
We, hence, assign the observed resonances to transitions from the
$1S_{3/2}$, $2S_{3/2}$, and $3S_{3/2}$ valence band multiplets to the
conduction band ground state. For $n_{\rm CdSe}=4$ and $5$, the
agreement with experimental data is comparable, albeit with a somewhat larger discrepancy between
the experimental and theoretical resonances ($\sim 20 \, {\rm nm}$).

In contrast to the resonance energies, the spectral weight of the
different resonances is not well reproduced by our theory.
Possible explanations are the failure of ${\bf k}\cdot {\bf p}$
theory, broken spherical symmetry, or a significant variation in
the $nS_{3/2}-1S_e$ exciton linewidth with $n$. For the
narrow quantum wells with $n_{\rm CdSe}=2-5$ studied here, first-principles
calculations may be more appropriate than ${\bf k}\cdot {\bf p}$ theory
for a rigorous description of the QDQW.
Broken spherical symmetry leads to a mixing of different valence
band multiplets. The resulting re-distribution of the spectral
weight from the $1S_{3/2}-1S_e$ transition to other exciton lines
decreases the spectral weight of the ground state exciton
transition.~\cite{meier:04b} In order to reduce the number of fit
parameters in Eq.~(\ref{eq:frot}), we have assumed that the
linewidths $\gamma_v$ of all $nS_{3/2}-1S_e$ exciton transitions
are identical. By allowing for a variation of $\gamma_v$ with $n$,
the agreement between experiment and theory in
Fig.~\ref{Fig1thy}(d) could be further improved.

In conclusion,  we have studied the spin dynamics and quantum size levels
in QDQWs using TRFR. The variation of the energy levels
and the electron $g$-factor with quantum well width allows one to selectively
address quantum wells using optical techniques. Possible future directions
include the investigation of the spin and orbital dynamics in more complex
heterostructures such as coupled quantum wells.

{\it Acknowledgments. --} This work was supported by ONR and
DARPA.


\begin{thebibliography}{35}
\bibitem[{rem()}]{rem2}
\bibinfo{note}{Low temperature measurements were performed on QDQW's embedded
  in an a polyvinyl butyral (PVB) matrix in a magneto-optical cryostat.}
\end{thebibliography}
\end{document}